\documentclass[preprint,review,12pt]{elsarticle}

\usepackage{amssymb}

\begin{document}

\begin{frontmatter}



\title{Anisotropic optical conductivity and electron-hole asymmetry in doped monolayer graphene in the presence of the Rashba coupling}


\author{S.S. Sadeghi}
\address{Department of Laser and
Optical Engineering University of Bonab, 5551761167 Bonab, Iran \\
Physics of  Department, Bu- Alisina University, Hamadan-Iran}
\author{A. Phirouznia\corref{cor1}}
\address{Department of Physics, Azarbaijan Shahid Madani University, 53714-161, Tabriz, Iran }
\cortext[cor1]{Corresponding author.} \ead{Phirouznia@azaruniv.ac.ir}
\author{V. Fallahi}
\address{Department of Laser and
Optical Engineering University of Bonab, 5551761167 Bonab, Iran.}

\begin{abstract}
In this study, the Optical conductivity of substitutionary doped
graphene is investigated in presence of the Rashba spin orbit
coupling (RSOC). Calculations have been performed within the
coherent potential approximation (CPA) beyond the Dirac cone
approximation. Results of the current study demonstrate that the
optical conductivity is increased by increasing the RSOC strength.
Meanwhile it was observed that the anisotropy of the band energy
results in a considerable anisotropic optical conductivity (AOC) in
monolayer graphene. The sign and magnitude of this anisotropic
conductivity was shown to be controlled by the external field
frequency. It was also shown that the Rashba interaction results in
electron-hole asymmetry in monolayer graphene.

\end{abstract}


\end{frontmatter}


\section{Introduction}
Recently graphene has attracted a rapidly growing interest due to
its potential application in nano-electronics \cite{1,2}. The
extremely high mobility of graphene, in comparison with conventional
semiconductors, constitutes a natural outstanding candidate to
design high performance electronic devices for next generation
electronics. Due to these surprising features of graphene, the
coming years has been named as carbon new age \cite{3}. However lack
of an energy gap in graphene is one of the biggest hurdles for
graphene
device applications \cite{4}.
\\
The effect of Spin orbit coupling (SOC) on the monolayer graphene
was found firstly by Kane and Mele \cite{5,6}. Strength of intrinsic
spin orbit coupling (ISOC) in comparison with the RSOC is very small
\cite{Rakyta,Gmitra,Huertas,Boettger,Konschuh,Abdelouahed}.
Meanwhile the RSOC strength in graphene has been reported up to
0.2eV \cite{Dedkov} where can be considered as a really high value
for a typical spin-orbit interaction. \\
The RSOC has been observed by impurity doping and external electric
field where the strength of doping induced RSOC is very weak
\cite{Ralph}. Meanwhile, it was shown that the RSOC can open and
control a noticeable gap at the Dirac points \cite{Gmitra,Qiao}.
\\
The effect of the RSOC on heavy doped graphene has been investigated
in \cite{13}. The effect of the impurities has been considered by
choosing chemical potential close to the M point in the first
Brillouin zone (BZ) \cite{13}. Results of this research show that
the RSOC in mono-layer graphene can strongly affect the optical
response of graphene \cite{13}.
\\
In the present study we formulated the effect of the impurities on
the optical conductivity of the mono-layer graphene in the presence
of RSOC. Numerical calculations have been performed under the
coherent potential approximation (CPA) where we have shown that the
density of states and the transition rate can be effectively
controlled by the density of impurities. It was shown that the RSOC
removes the electron-hole band symmetry. In this way we have to
consider the deformation of the band energies and density of states
as result of the impurities.
\\
Owning to the fact that the RSOC Hamiltonian is translationally
invariant, it can be formulated under CPA approach as shown in later
sections. The CPA is one of the important and widely used method for
describing disordered systems. The problem is that how conductivity
of disordered graphene can be affected by the RSOC under CPA
approach?
\\
In this manuscript, band structure of graphene has been studied
beyond the Dirac cone approximation in framework of tight-binding
Hamiltonian. As a result, trigonal warping (TW) of the energy bands
can contribute in optical conductivity. The TW is observed when
Fermi circle of a degeneracy point lead to deformation \cite{14} by
increasing the Fermi energy. TW was expected to be responsible for
anisotropic optical conductivity (AOC) as it removes isotropic band
cross sections at Fermi level, however the results of the current
study show that, even at low Fermi energies i.e. when the TW is
negligible the amount of the AOC is really significant. Dirac-cone
approximation suppresses the TW effect and therefore the intrinsic
anisotropy of the bands (regardless of the position of the Fermi
energy) can be responsible for AOC of single layer graphene. It
seems that the TW could produce some kind of anisotropic effects
when, only the occupied states are allowed to contribute in this
process. Meanwhile it should be noted that in some of the physical
processes such as optical absorption, where a transition is required
between an occupied and another empty state. The anisotropy of the
entire band energies could result in AOC. The effect of the RSOC on
the AOC of a monolayer graphene has also been studied in the current
work. The AOC manifests itself when the the optical conductivity was
different along $x$ and $y$ directions.
\\
Electron density of state (DOS), the optical conductivity of
graphene can be controlled by manipulating gate voltage which
controlls the RSOC strength. Optical properties of a pure monolayer
graphene has been studied in \cite{15}. Where it was shown that RSOC
can result in a controllable blue shift. In the present work,
Density of state, AOC of disordered graphene calculated beyond the
Dirac-cone approximation.
\section{Theoretical model}
\begin{figure}[htbp]
\centering
  \includegraphics{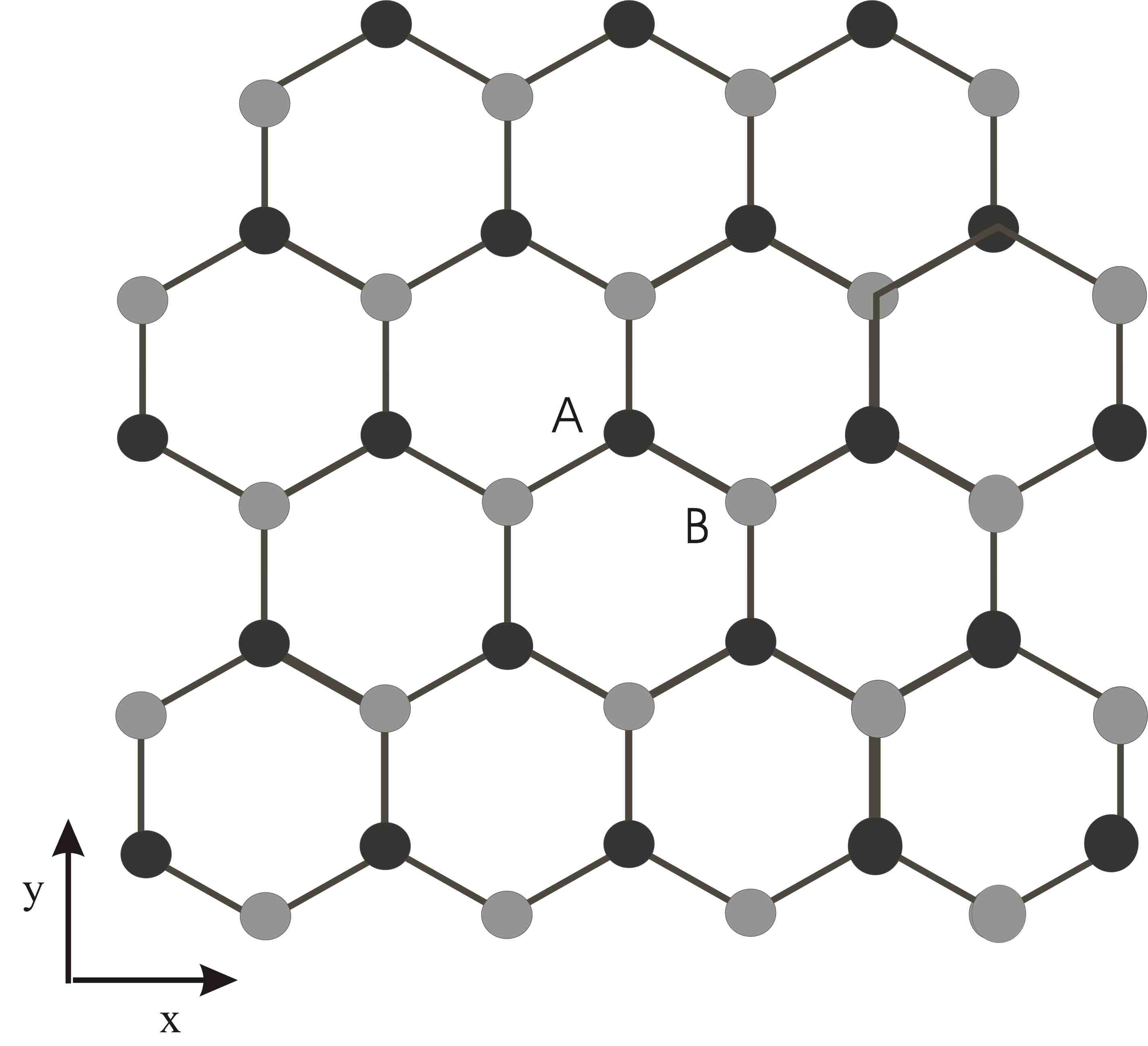}
  \caption{Spatial orientation of the monolayer graphene with respect to the $x$ and $y$
  axis and A, B sublattices.} \label{fig1}
\end{figure}

In the current work we have considered a monolayer graphene, in
which its two-dimensional lattice has been oriented with respect to
the $x$ and $y$ axis as shown in Fig. \ref{fig1}.
\\
Nearest neighbor-tight binding Hamiltonian of pure graphene reads
\begin{eqnarray}
H_0=-t\sum_{<i,j>,\sigma }\,({{a}^{\dagger }}_{\sigma
,i}{{b}_{\sigma ,j}}+h.c)
\end{eqnarray}
Where the operators $a_{\sigma ,i}^{\dagger }$ and ${{b}_{\sigma
,j}}$ denote the creation and annihilation of an electron with spin
${\sigma}$ in sublattices A and B (Fig. \ref{fig1}), respectively.
Where $t=2.66$ eV is hopping parameter.
\\
RSOC can generate a gap in graphene and converts graphene to
semiconductor. According to the results of density functional theory
and other approaches, the strength of the ISOC is several orders of
magnitude weaker than the strength of RSOC (the strength of ISOC is
about 1-50$\mu$eV)
\cite{Rakyta,Gmitra,Huertas,Boettger,Konschuh,Abdelouahed,Dedkov}.
Therefore in this paper the effect of ISOC ( also ISOC in next
nearest neighbor) is neglected. \\
RSOC in graphene and other structures emerges when lattice inversion
symmetry is broken \cite{Rashba}. In this study RSOC can be
considered to be induced by perpendicular gate voltage or by
coupling with the substrate. Rashba coupling is given in the form of
a nearest neighbor hopping term as follows \cite{Ralph}
\begin{eqnarray}
{{H}_{R}}=-{{t}_{R}}\sum_{<i,j>}\,c_{i}^{\dagger }(\textbf{s}\times
\hat{{{d}_{{ij}}}}).{\mathop{{\hat{z}}{{{c}}_{j}}}}\,+h.c.
\end{eqnarray}
$\textbf{s}$ is the vector of Pauli matrices, $\hat{{{d}_{{ij}}}}$
is the unit vector that connects the i and j lattice sites and $t_R$
is the strength of RSOC where $<i,j>$ indicates that the sum is
performed over the nearest neighbors (Fig. \ref{fig1}).
\\
Matrix representation of total Hamiltonian on wave function $\psi
=({{\psi }_{A\uparrow }},{{\psi }_{A\downarrow }},
   {{\psi }_{B\uparrow }}, {{\psi }_{B\downarrow }})$ is given by
\\
\begin{eqnarray}
\,{{H}_{0}}=\left(
\begin{array}{llll}
 0 & 0 & \gamma(\textbf{k})   & 0 \\
 0 & 0 & 0  & \gamma(\textbf{k})  \\
 \gamma^*(\textbf{k})    & 0  & 0 & 0 \\
 0 & \gamma^*(\textbf{k})   & 0 & 0
\end{array}
\right),
\end{eqnarray}
and
\begin{eqnarray}
\,{{H}_{R}}=\left(
\begin{array}{llll}
 0 & 0 & 0   & \beta_+(\textbf{k}) \\
 0 & 0 & \beta_-(\textbf{k})  & 0  \\
 0 & \beta_-^*(\textbf{k})  & 0 & 0 \\
 \beta_+^*(\textbf{k}) & 0  & 0 & 0
\end{array}
\right).
\end{eqnarray}
where $\gamma(\textbf{k})
={{e}^{-ia{{k}_{y}}}}+2{{e}^{-i\frac{a}{2}{{k}_{y}}}}\cos
({{a}_{1}}{{k}_{x}})$ ${{\beta }_{\pm }}=i\,{{t}_{R}}({{\xi
}_{1}}(k)\pm {{\xi }_{2}}(k))$, $ {{\xi
}_{1}}(k)={{e}^{i{{a}_{1}}{{k}_{x}}}}({{e}^{-i{{a}_{2}}{{k}_{y}}}}-\cos
({{a}_{1}}{{k}_{x}}))$ and $ {{\xi
}_{2}}(k)=\sqrt{3}{{e}^{i{{a}_{1}}{{k}_{x}}}}\sin
({{a}_{1}}{{k}_{x}})$ in which
${{a}_{1}}=\frac{\sqrt{3}a}{2},\,{{a}_{2}}=\frac{3}{2}a$ in which
the carbon-carbon distance is denoted by $a=1.42${\AA}.
\\
Using the perturbation theory we can obtain the eigenstates of
$H_0+H_R$ as follows
\begin{eqnarray}
\left| {1k} \right\rangle =\left(
\begin{array}{l}
   1 \\
  0 \\
  -\varphi_k^{*} \\
  \varphi_k^{*} \\
\end{array} \right),~~~
\left| {2k} \right\rangle =\left( \begin{array}{l}
  0 \\
  1 \\
  \varphi_k^{*} \\
  -\varphi_k^{*} \\
\end{array} \right), \\
\left| {3k} \right\rangle =\left( \begin{array}{l}
  -{\varphi_k} \\
  {\varphi_k} \\
  1 \\
  0 \\
\end{array} \right),~~~
\left| {4k} \right\rangle =\left( \begin{array}{l}
  {\varphi_k} \\
  -{\varphi_k} \\
  0 \\
  1 \\
\end{array} \right),
\end{eqnarray}
where we have defined $\varphi_k ={{x}_{k}}/({4{{\left| \gamma_k
\right|}^{2}}}),\,\,\,\,{{x}_{k}}=\gamma_k \beta _{-}^{*}-{{\gamma_k
}^{*}}{{\beta }_{+}}$.
\\
Where each state corresponds to the following eigenvalues
\begin{eqnarray}
 E_{1,k}=-\left| \gamma_k  \right|-4\left| \gamma_k  \right|\,{{\left| \varphi_k  \right|}^{2}}
\,\,\,,\,\,\,E_{2,k}=-\left| \gamma_k  \right|-4\left| \gamma_k  \right|\,{{\left| \varphi_k  \right|}^{2}} \\
E_{3,k}=\left| \gamma_k  \right|+4\left| \gamma_k \right|\,{{\left|
\varphi_k \right|}^{2}} \,\,\,\,,\,\,\,\,E_{4,k}=\left| \gamma_k
\right|+4\left| \gamma_k \right|\,{{\left| \varphi_k \right|}^{2}}
\end{eqnarray}
It should be noted that the exact expression for these eigenstates
and eigenvalues are also available however, we have observed that
the use of these compact form of the eigenstates and eigenvalues,
given by the perturbation theory, results in a significant time
saving during the self consistent computations.
\\
Under of CPA theory Hamiltonian of electron in this system is $H=H
_0+H_R+V$ which $V$ is the short range potential arising from the
substitutional impurities. The Green's function of pure system,
${G}_{0}$, is given by
\begin{eqnarray}
{{G}_{0}}(z)=\frac{1}{z-{{H}_{0}}-H_R}
\end{eqnarray}
In the CPA approach the Hamiltonian is given by
\begin{eqnarray}
H=H_0+H_R+\Sigma+V-\Sigma
\end{eqnarray}
The site dependent self energies $\Sigma(E)$ should be determined
during the CPA algorithm. By using a self-consistent approach one
can obtain as much as possible accurate self-energies using the
following relations
\begin{eqnarray}
\label{11} G_{eff}={{G}}+{{G}}V' G_{eff}.
\end{eqnarray}
In which we have defined
\begin{eqnarray}
V'&=&V-\Sigma(E) \\
{{G}}(z)&=&\frac{1}{z-{{H_{eff}}}}
\\
H_{eff}(E)&=&H_0+H_R+\Sigma(E).
\end{eqnarray}
Equation (\ref{11}) can be restated in term of the $T$-Matrix
formalism as follows
\begin{eqnarray}
\label{15} G_{eff}&=&{{G}}+{{G}}T {{G}},
\\
T&=&V'+V' GT.
\end{eqnarray}
The effective self energies can be determined if the following
condition is satisfied
\begin{eqnarray}
\label{16} <G_{eff}>={{G}},
\end{eqnarray}
i.e. when
\begin{eqnarray}
\label{17} <T>=0.
\end{eqnarray}
In which $<>$ denotes the configurational averaging.
\\
When we applied self-consistency to average T-matrix, finally site
independent self-energy of effective medium achieved. The CPA
self-consistent equations in the unit cell may be restated as
\begin{eqnarray}
\Sigma_{\lambda, A/B}(z) =(x/2){{\varepsilon
}_{A/B}}+(1-x/2){{\varepsilon }_{A/B}}-({{\varepsilon
}_{A/B}}-\Sigma_\lambda ){{\bar{G}}^\lambda_{00}}({{\varepsilon
}_{A/B}}-\Sigma_\lambda )
\end{eqnarray}
Which  ${{\bar{G}}_{00}}(z)$  trace of all configurations in real
space.
\begin{eqnarray}
\label{20} {{\bar{G}}^\lambda_{00}}(z)=\frac{1}{{{\Omega
}_{1BZ}}}\int\limits_{1BZ}{\frac{d{{k}_{x}}d{{k}_{y}}}{(z-E_\lambda(k)-\Sigma_\lambda
)}},
\end{eqnarray}
this integration is over the entire range of the first Brillouin
zone. Therefore using the Dirac point approximation in
$E_\lambda(k)$ gives incorrect results. Consequently the Dirac point
approximation cannot be employed in the CPA approach. In the above
expressions $z=E+i\epsilon$ where $\epsilon$ is infinitesimally
small positive number. $A$, $B$ referred to the different
sublattices of the graphene and $x$ is the density of impurities.
Since $A$ and $B$ sublattices can be treated on the same footing we
can write $\Sigma_{\lambda, A}=\Sigma_{\lambda, B}$ and the density
of impurities on each sublattice will be identical and therefore
given by $x/2$.
\\
We have employed a diagonal self energy matrix as
follows \cite{Nilsson}
\begin{eqnarray}
\,{{\Sigma}}=\left(
\begin{array}{llll}
 \Sigma_1(E) & 0 & 0   & 0 \\
 0 & \Sigma_2(E) & 0  & 0  \\
 0    & 0  & \Sigma_3(E)& 0 \\
 0 & 0   & 0 & \Sigma_4(E)
\end{array}
\right),
\end{eqnarray}
When the self energies is finally obtained. Density of state in term
of Green's function is expressed as
\begin{eqnarray}
{D(E)}=-\frac{1}{\pi }{Tr Im}<G(E)>
\end{eqnarray}
\section{Optical conductivity}
Real part of conductivity in Linear response theory for disordered
systems at arbitrary incident energy, $\omega$, given in term of
Kobu-GreenWood equation \cite{30,31}
\begin{eqnarray}
\sigma_{1x_ix_j}\left( \omega  \right)=\frac{e^{2}\hbar}{S \pi}\int
\frac{f\left( E \right)-f\left( E+\hbar \omega\ \right)}{\hbar
\omega}d E~ Tr\{{{v}_{x_i}}{Im G}(E+\hbar \omega){{v}_{x_j}}Im
G(E)\}
\end{eqnarray}
$f(E)$ is Fermi-Dirac distribution, $v_{x_i}$ is velocity of
electron in $x_i$ direction and $S$ is surface of our system we have
${{v}_{x_i}}=dx_i/dt=(i/\hbar)[H_0+H_R,\,x_i]$. Since $G_{eff}(E)$
is diagonal in k-space representation, vertex correction cannot be
captured in this method. We can write
\begin{eqnarray}
\sigma_{1x_ix_j}(\omega )&=&\frac{\hbar {{e}^{2}}}{\pi \Omega
}\int{dE
\frac{f(E)-f(E+\hbar \omega )}{\hbar \omega }}\\
&&\times\sum_{\eta {\eta }'}\sum_{k{k}'} \left| \left\langle  \eta k
| \left. {{v}_{x}} \right|{\eta }'{k}' \right\rangle
\right|^{2}{Im}{{G}_{{\eta }'{k}'}}(E+\hbar \omega ){Im}{{G}_{\eta
k}}(E).\nonumber
\end{eqnarray}
In which
\begin{eqnarray}
\left\langle  \eta k | \left. {{v}_{x}}\, \right|{\eta }'{k}'
\right\rangle &=&\frac{i}{S\hbar }
\int{{{e}^{i({k}'-k).\vec{r}}}\,x\,dr}({{E}_{\eta k}}-{{E}_{{\eta
}'{k}'}})\left\langle  \eta  | {{\eta }'} \right\rangle \\ {{\left|
\left\langle  \eta k | \left. {{v}_{x}}\, \right|{\eta }'{k}'
\right\rangle  \right|}^{2}}&=&\frac{1}{{{\hbar
}^{2}}}\frac{{{({{E}_{\eta k}}-{{E}_{{\eta }'{k}'}})}^{2}}}{{{\left|
{{{{k}'}}_{x}}-{{k}_{x}} \right|}^{2}}}{{\delta
}_{{{k}_{y}}\,{{{{k}'}}_{y}}}}{{\left| \left\langle  {{\eta }'} |
\eta  \right\rangle  \right|}^{2}}.
\end{eqnarray}
Similarly imaginary part of the Green's function is given by
\begin{eqnarray}
{Im}\,{{G}_{\eta k}}(E)=\frac{{Im}\,{{\Sigma }_{\eta
}}}{{{(E-{{E}_{\eta k}}-{Re}\,{{\Sigma }_{\eta
}})}^{2}}+{{({Im}\,{{\Sigma }_{\eta }})}^{2}}}
\end{eqnarray}
Finally the AOC can be obtained by
\begin{eqnarray}
AOC=\frac{\sigma_{xx}-\sigma_{yy}}{\sigma_{xx}+\sigma_{yy}}
\end{eqnarray}
\section{Results and discussions}
\begin{figure}[htbp]
\centering
  \includegraphics[width=5in]{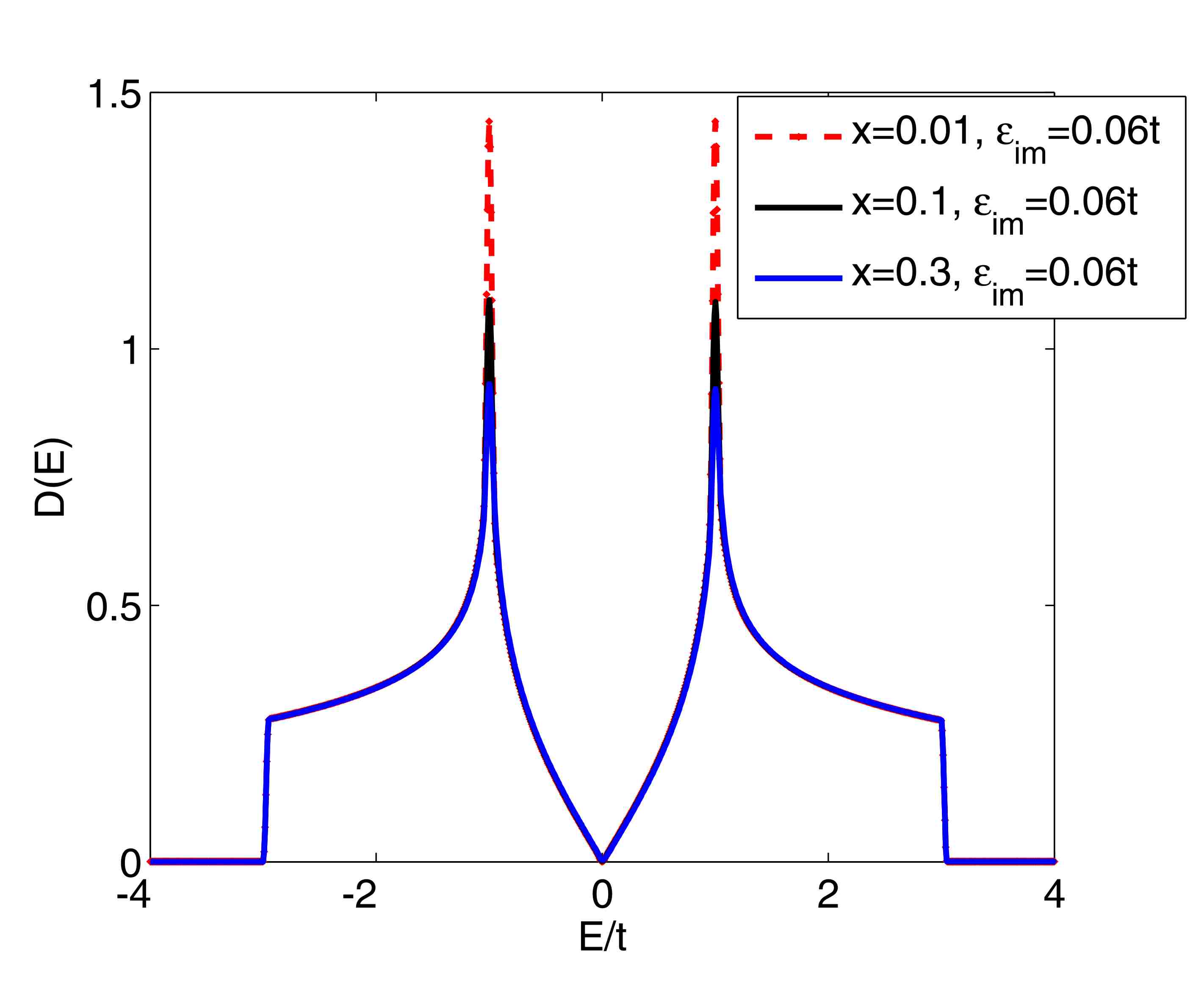}
  \caption{Density of states in graphene at different impurity densities and zero Rashba coupling.}
\label{fig2}
\end{figure}

\begin{figure}[htbp]
\centering
  \includegraphics[width=5in]{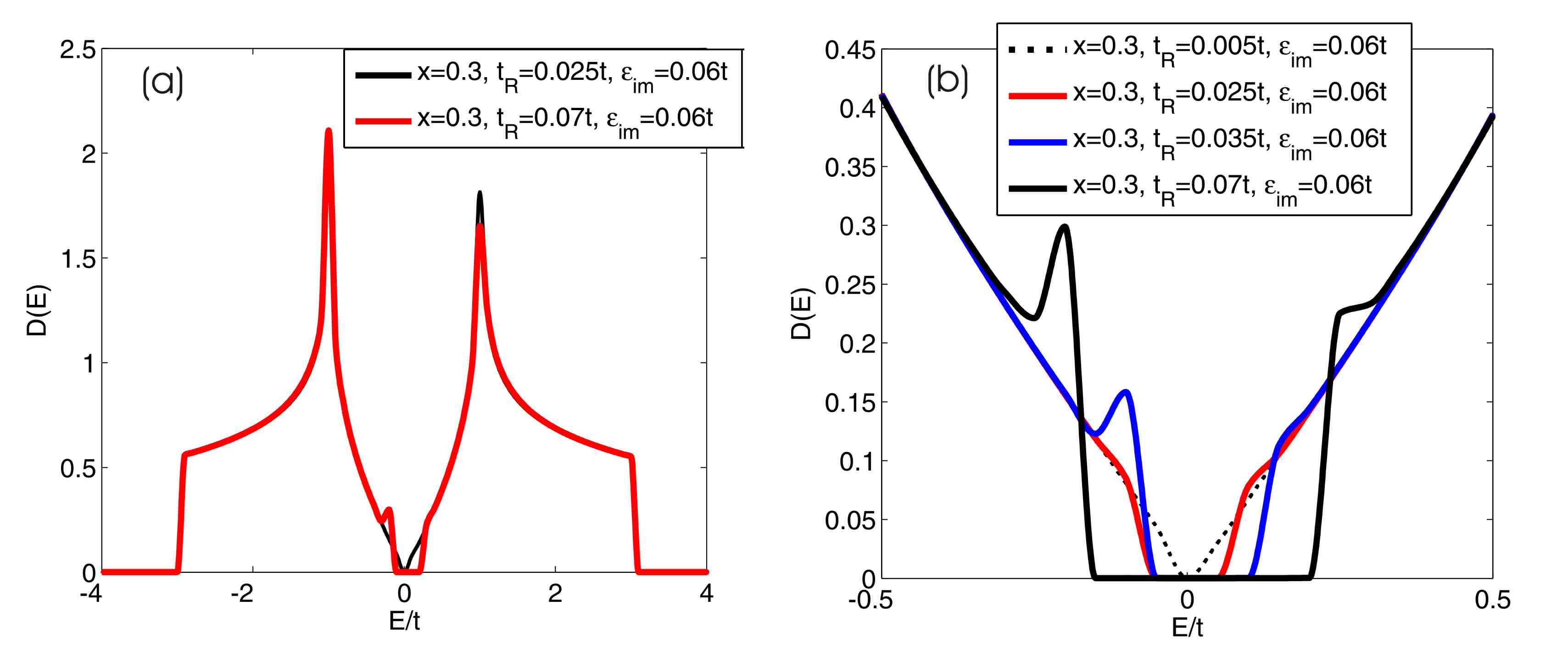}
  \caption{Density of states in graphene at different Rashba couplings.
  The symmetry of the conduction ($E>0$) and valence ($E<0$) bands has been broken by the Rashba interaction.}
\label{fig3}
\end{figure}

\begin{figure}[htbp]
\centering
  \includegraphics{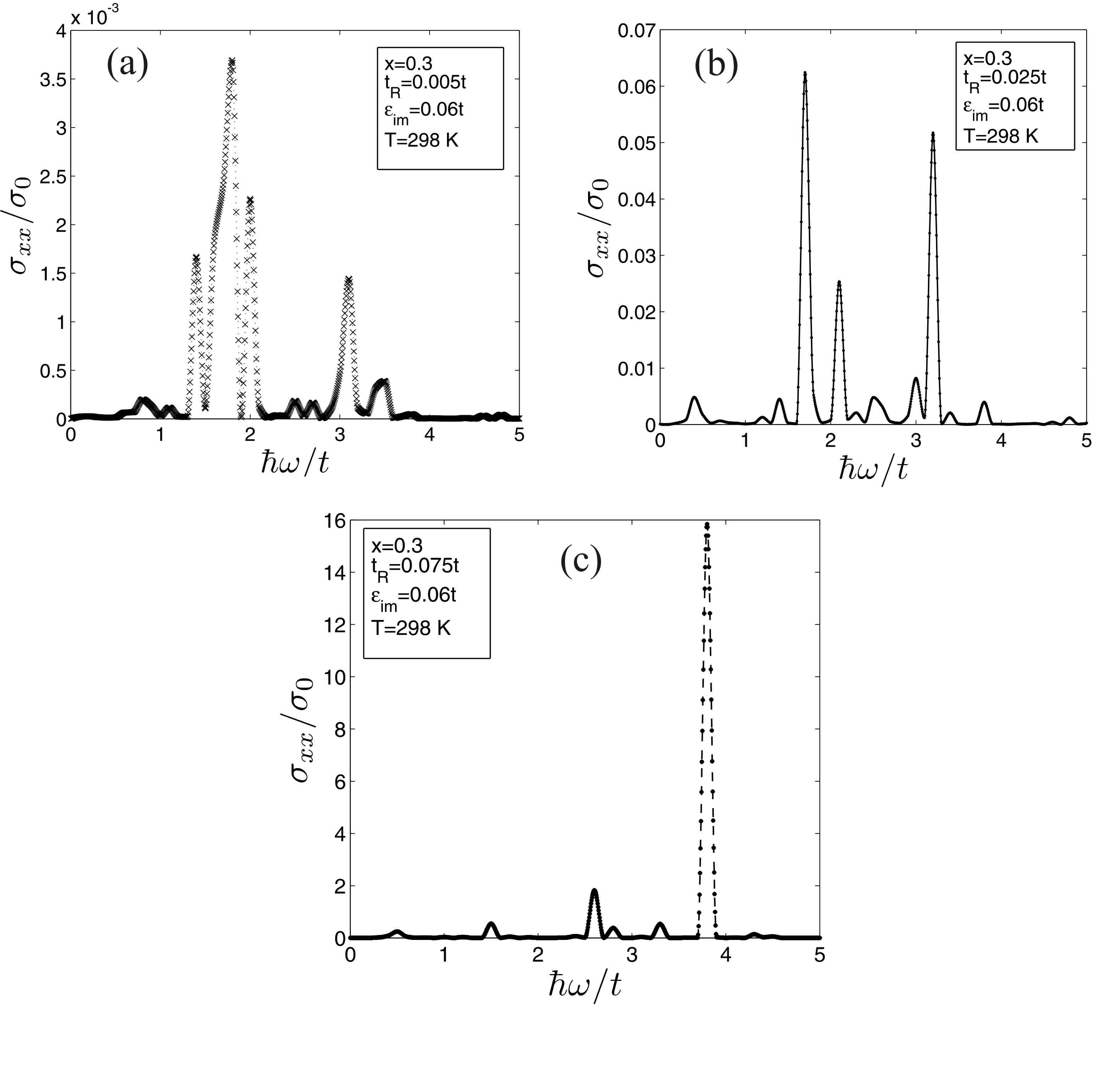}
  \caption{ Optical conductivity along the x axis at different Rashba couplings (${{\sigma
}_{0}}=\frac{{{e}^{2}}}{\hbar }$ ).} \label{fig4}
\end{figure}

\begin{figure}[htbp]
\centering
  \includegraphics{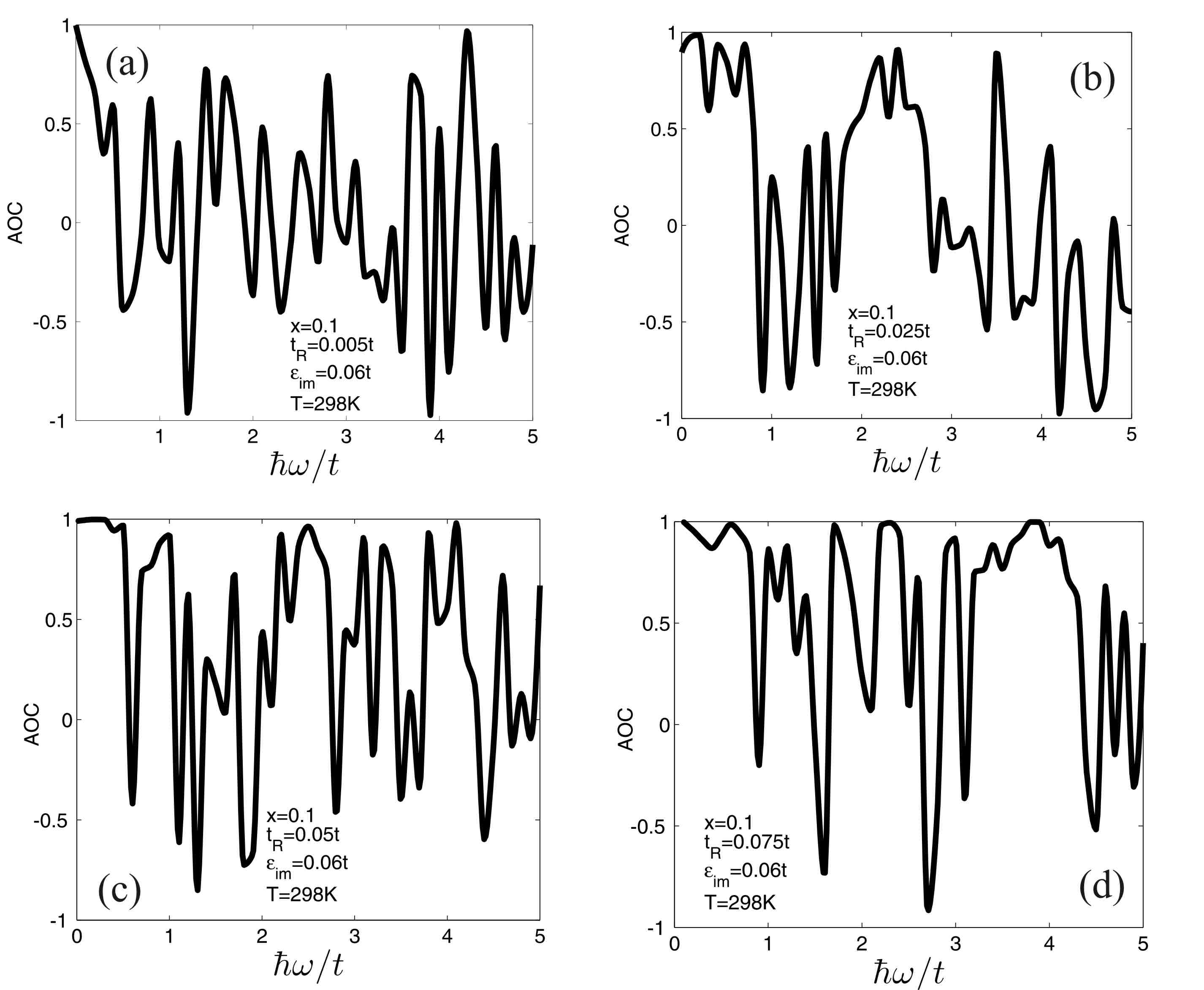}
  \caption{Anisotropic optical conductivity as a function of the photon energy at different Rashba couplings.}
\label{fig5}
\end{figure}

The results of the current study show that substrate induced
spin-orbit coupling could change the density of states and the gap
energy, introduced by the impurities Figs \ref{fig2}-\ref{fig3}.
Therefore presence of the impurities in the graphene, results in
some additional effects other than the change in Fermi energy.
Meanwhile it should be noted that increasing the Rashba coupling
removes the electron-hole symmetry as shown in the Fig. \ref{fig3}.
It can be inferred by comparison of the Figs \ref{fig2}-\ref{fig3}
that in the absence of the Rashba interaction, single layer graphene
has a symmetric density of states in both $E>0$ and $E<0$ ranges of
energy and substitutional impurities can not remove this symmetry,
however when we switch on the Rashba coupling a significant electron
hole asymmetry arises Figs \ref{fig2}-\ref{fig3}.
\\
The effect of the RSOC on the optical conductivity of doped graphene
has been shown for different Rashba coupling strengths Fig.
\ref{fig4}. Since splitting of the band energies is effectively
increased by Rashba coupling the blue shift of the absorption curve
is expected by increasing the Rashba coupling. These results
confirms that, increasing the Rashba coupling strength,
simultaneously increases and shifts the real part of the optical
conductivity in doped graphene. This is in agreement with the
results of the Rashba coupling induced blue shift in the pure
graphene.
\\
Another interesting feature of the obtained results is the
considerable AOC in the monolayer graphene as shown Fig. \ref{fig5}.
Numerical results show that the optical conductivity (and therefore
optical absorption)is highly dependent on the direction of the
external field polarization, where the sign and amount of this
anisotropy is determined by the frequency of the external field
(Fig. \ref{fig5}). The RSOC can slightly modify the AOC of the doped
graphene at any given frequency, however AOC mainly depends on our
approach which has been performed beyond the Dirac point
approximation.
\\
At the level of the Dirac point approximation we would get a fully
symmetric circle shaped of the Fermi surfaces centered at each
Brillouin zone corner. Since the absorption process and any
scattering process could take place mainly between the occupied and
empty states. Therefore the whole Brillouin zone could contribute
into the photon absorption. Accordingly the absorption was not
limited to the symmetric parts of the occupied Brillouin zone. The
oscillation which has been observed in the AOC as a function of the
frequency (Fig. \ref{fig5}) can be explained by considering the
fact, that the imaginary part of the Green functions actually
guarantee the energy conservation within a finite broadening range.
These conservation rules play a selective effect in the absorption
and therefore the AOC cannot be a monotonic function of the photon
energy.
\\
As mentioned before the AOC is slightly affected by the Rashba
interaction. Meanwhile it should be noted that the Rashba coupling
itself is not invariant under the $x \leftrightarrow y$ interchange
and therefore it was expected that this interaction should change
the AOC of the monolayer graphene. However since the Rashba
interaction is really small in comparison with the bare graphene
Hamiltonian and the change of the AOC by the Rashba coupling will be
quite limited.
\section{Conclusion}
In the present work we have obtained the effect of the Rashba
coupling on optical conductivity in doped graphene. Results of this
work show that the optical conductivity of the graphene is
essentially anisotropic. We have also shown that the Rashba
interaction removes the electron-hole symmetry in monolayer
graphene.







\newpage
Fig. 1
\\
Spatial orientation of the monolayer graphene with respect to the
$x$ and $y$ axis and A, B sublattices.
\\
Fig. 2
\\
Density of states in graphene at different impurity densities and
zero Rashba coupling.
\\
Fig. 3 \\
Density of states in graphene at different Rashba couplings. The
symmetry of the conduction ($E>0$) and valence ($E<0$) bands has
been broken by the Rashba interaction.
\\
Fig. 4
\\
Optical conductivity along the x axis at different Rashba couplings
(${{\sigma }_{0}}=\frac{{{e}^{2}}}{\hbar }$ ). \\
Fig. 5
\\
Anisotropic optical conductivity as a function of the photon energy
at different Rashba couplings.
\newpage

\end{document}